\documentclass[10pt,twocolumn,a4paper]{article}

\usepackage[left=15mm, right=15mm, top=15mm, bottom=15mm]{geometry}

\usepackage{subfig}
\usepackage{flushend}
\usepackage{indentfirst}
\usepackage{graphics}
\usepackage{amsmath}
\usepackage{graphicx}
\usepackage{epstopdf}
\usepackage{float}
\usepackage{amsthm}
\usepackage[font=footnotesize,labelfont=bf]{caption}
\usepackage[labelsep=period]{caption}

\theoremstyle{plain}
\newtheorem{remark}{Remark}
\graphicspath{{./figure/}}

\begin{document}

\title{\huge \textbf{Inference in the stochastic Cox-Ingersol-Ross diffusion process with continuous sampling: Computational aspects and simulation}}

\twocolumn[
\begin{@twocolumnfalse}

\author{\textbf{Ahmed Nafidi}$^{1}$, \textbf{Abdenbi El Azri}$^{1,*}$\\\\
\footnotesize $^{1}${Hassan First University of Settat, National School of Applied Sciences Berrechid, LAMSAD, B.P. 218, 26103 Berrechid, Morocco.}\\
\footnotesize $^{*}$Corresponding Author: a.elazri@uhp.ac.ma}

\date{}

\maketitle

\end{@twocolumnfalse}
]

\noindent \textbf{\large{Abstract}} \hspace{2pt} In this paper, we consider a stochastic model based on the Cox- Ingersoll- Ross model
(CIR). The stochastic model is parameterized analytically by applying Itô’s calculus and the trend functions of the proposed process is calculated. The parameter estimators are then derived by means of two procedures : the first is used to estimate the parameters in the drift coefficient by the maximum likelihood (ML) method, based on continuous sampling, and the second procedure approximates the diffusion coefficient by two methods. Finally, a simulation of the process is presented. Thus, a typical simulated trajectory of the process and its estimators is obtained.\\

\noindent \textbf{\large{Keywords}} \hspace{2pt} Cox-Ingersoll-Ross model, Stochastic diffusion process, Stochastic differential equation, Maximum likelihood method, Simulation.\\

\noindent\hrulefill
\section{\Large{Introduction}}
 In the present paper, we consider the Cox- Ingersoll- Ross model (or CIR model).
The CIR model describes the evolution of interest rates. It was introduced by John Carrington Cox, Jonathan Edwards Ingersoll and Stephen Alan Ross in 1985 (cf.\cite{ref1}). It has been applied in finance for to describes the evolution of interest rates. Moreover, it is used by Heston, S. L. \cite{ref6}, for the stochastic volatility model and by Duffie, D. \cite{ref4}, for the default intensities in credit risk model. Lingjiong Zhu \cite{ref11}, proposed a generalization of the classical Cox-Ingersoll-Ross process and the classical Hawkes process with exponential exciting function. S. Dyrting \cite{ref12} tested the existing methods for evaluating the noncentral $\chi^2-$ditribution for the Cox-Ingersoll-Ross process and  developed a new method based on a Bessel series representation. M. A. Jafari, S. Abbasian \cite{ref5}, determined the moments for solution of the Cox-Ingersoll-Ross Interest Rate Model. Kalana Nuwanpriya Alupotha \cite{ref2}), given the derivation of probability density function of CIR
model under a specific condition.

This paper is organized as follows: in the second section, the parametrization and the transition probability density function and the mean value of the CIR diffusion process are obtained. In the third, the estimators of the parameters in the drift coefficient are derived by the maximum likelihood method,  based on continuous sampling and the diffusion coefficient estimator is approximated. In the fourth, presents the results obtained from the simulation examples and its parameter estimators. Finally, the briefly summarizes and concludes from this study.

\section{\Large{The model and its characteristics}} \label{sec021}
\subsection{\normalsize \textbf{The model}}
The stochastic model proposed is based on the CIR model, which is
defined as a diffusion process $\{x(t) ;t \in [0;T]\}$, with values in $(0;\infty)$, and sample paths that are almost surely continuous
and with infinitesimal moments (drift and diffusion coefficient) that are given by
$$ a(t,x)=\kappa (\theta- x);\qquad b(t,x)=\sigma^{2}x$$
where $\kappa $ is the mean reversion speed, $\theta$ is mean reversion parameter, and $\sigma$ Standard deviation that determines the volatility and $\kappa $,$\theta$ and $\sigma$ are real parameters (to be estimated). So, we consider the following SDE:
\begin{equation}\label{p2}
    dx(t)=\kappa \left(\theta- x(t)\right)dt+\sigma\sqrt{x(t)}dw(t);\quad x(0)=x_{0}
\end{equation}

 An examination of the boundary classification criteria shows that $x(t)$ can reach zero if $\sigma^{2}>2\kappa \theta$. If $\sigma^{2}\leq 2\kappa \theta$, the upward drift is sufficiently large to make the origin inaccessible. In either case, the singularity of the diffusion coefficient at the origin implies that an initially nonnegative interest rate can never subsequently become negative. Considering the analytical properties of $a(t, x)$ and $b(t, x)$, it follows that the SDE (\ref{p2}) has a unique solution $\{x(t), t \in [0, T]\}$ which is a $(0,\infty)$ valued diffusion process with an initial
value $x_0,$ (cf.\cite{ref13}).

 Let $\alpha=\kappa \theta$ and $\beta=\kappa $. After substitution in equation (\ref{p2}), we obtain the following SDE:
 \begin{equation}\label{p3}
    dx(t)=\left(\alpha-\beta x(t)\right)dt+\sigma\sqrt{x(t)}dw(t);\quad x(0)=x_{0}
\end{equation}

\subsection{\normalsize \textbf{A parametrization of the CIR model}}\label{sec022}
By means of the appropriate transformation of the form $y(t) = e^{\beta t}x(t)$ , and by using
the Itô rule, the SDE (\ref{p3}) becomes
$$d\left(e^{\beta t}x(t)\right)=\alpha e^{\beta t} dt+\sigma e^{\beta t}\sqrt{x(t)}dw(t)$$
 by integrating both sides yields,
 \begin{equation*}
    e^{\beta t}x(t)=x_{s} e^{\beta s}+\frac{\alpha}{\beta}\left(e^{\beta t}-e^{\beta s}\right)+\sigma\int _{s}^{t}e^{\beta \tau}\sqrt{x(\tau)}dw(\tau)
\end{equation*}
Finally, we have:
\begin{equation}\label{p4}
\begin{split}
   x(t)&=x_{s} e^{\beta (s-t)}+\frac{\alpha}{\beta}\left(1-e^{\beta (s-t)}\right)  \\
    & +\sigma\int _{s}^{t}e^{\beta (\tau-t)}\sqrt{x(\tau)}dw(\tau).
\end{split}
\end{equation}

If $s=0$, we deduce that the parametrization of the original SDE (\ref{p3}) is:
\begin{equation}\label{p5}
    x(t)=x_{0} e^{-\beta t}+\frac{\alpha}{\beta}\left(1-e^{-\beta t}\right)+\sigma\int _{0}^{t}e^{\beta (\tau-t)}\sqrt{x(\tau)}dw(\tau)
\end{equation}

\subsection{\normalsize \textbf{Transition probability density function of the model}}\label{sec023}
  The transition probability density function of the CIR process $x(t)$, given $x(s)$ for $s < t$. In fact, the transition law of
$x(t)$ is a non-central chi-square distribution:
$$x(t)|x(s)\sim \zeta\chi ^2_k(\lambda)$$
with degrees of freedom $k=\frac{4\alpha}{\sigma^2}$, $\zeta=\frac{\sigma^2\left(1-e^{-\beta(t-s)}\right)}{4\beta}$ and non-centrality parameter
 $ \lambda =\frac{4\beta e^{-\beta(t-s)}x(s)}{\sigma^2\left(1-e^{-\beta(t-s)}\right)}$.
Hence, the expression for the transition probability density function as obtained by John C. Cox and al. \cite{ref1} of the model is

\begin{align*}
  f\left(x,t|y,s\right)=&\frac{2\beta}{\sigma^{2}\left(1-e^{-\beta(t-s)}\right)}\\
  \times&\exp\left(\frac{-2\beta\left(x+ye^{-\beta(t-s)}\right)}{\sigma^{2}\left(1-e^{-\beta(t-s)}\right)}\right)\left(\frac{x}{ye^{-\beta(t-s)}}\right)^{\frac{q}{2}}\\
 \times & I_{q}\left(\frac{4\beta\sqrt{xy}e^{\frac{-\beta(t-s)}{2}}}{\sigma^{2}\left(1-e^{-\beta(t-s)}\right)}\right),
\end{align*}
where $I_{q}$ denotes the modified Bessel function of the first kind and $q=\frac{2\alpha}{\sigma^{2}}-1$.

\subsection{\normalsize \textbf{Computation of the trend function}}\label{sec024}
Using the expression (\ref{p5}), we deduce the conditional trend function (CTF) of the process is
\begin{align*}
  E(x(t)|x(s)=x_{s})&=E\left[ x_{s} e^{\beta (s-t)}+\frac{\alpha}{\beta}\left(1-e^{\beta (s-t)}\right)\right. \\
   & +\left.\sigma\int _{s}^{t}e^{\beta (\tau-t)}\sqrt{x(\tau)}dw(\tau)\right].
\end{align*}

Then, we have
\begin{align*}
  E(x(t)|x(s)=x_{s}) &= x_{s} e^{\beta (s-t)}+\frac{\alpha}{\beta}\left(1-e^{\beta (s-t)}\right)  \\
   & +E\left(\sigma\int _{s}^{t}e^{\beta (\tau-t)}\sqrt{x(\tau)}dw(\tau)\right).
\end{align*}

The random variable in the last expression is normally distributed with mean zero, then the final form of the mean value of CIR is
\begin{equation}\label{CTF}
 E(x(t)|x(s)=x_{s})= x_{s} e^{\beta (s-t)}+\frac{\alpha}{\beta}\left(1-e^{\beta (s-t)}\right).
\end{equation}
Finnaly, if $s=0$, the trend function (TF) of CIR (mean value of CIR) is given by the following expression:
\begin{equation}\label{TF}
E(X_{t})= x_{0} e^{-\beta t}+\frac{\alpha}{\beta}\left(1-e^{-\beta t}\right).
\end{equation}

\begin{remark}
 {\normalfont Note that, if $\beta>0$, thus
$$\lim\limits_{t\rightarrow \infty} E(X_{t})=\frac{\alpha}{\beta}.$$}
\end{remark}

\subsection{\normalsize \textbf{A confidence interval of the CIR}}
Let $v(s,t)=x(t)|x(s)=x_s$. We known that $v(s,t)\sim \zeta\chi ^2_k(\lambda)$, with degrees of freedom $k=\frac{4\alpha}{\sigma^2}$, $\zeta=\frac{\sigma^2\left(1-e^{-\beta(t-s)}\right)}{4\beta}$ and non-centrality parameter
 $\lambda =\frac{4\beta e^{-\beta(t-s)}x(s)}{\sigma^2\left(1-e^{-\beta(t-s)}\right)}$. Therefore, the random variable
 $\frac{v(s,t)}{\zeta} \sim \chi ^2_k(\lambda)$.
 We make the approximation of the chi-square by the standard normal distribution \cite{ref17}. So, the random variable $z$ is given by
$$z=\frac{\frac{v(s,t)}{\zeta}-(k+\lambda)}{\sqrt{2(k+2\lambda)}} \sim N(0,1),$$ when $k\rightarrow \infty$ or $\lambda\rightarrow \infty$.

A $100(1-\alpha) \%$ conditional confidence interval for $z$ is given by $P(-\xi\leq z\leq \xi)=1-\alpha$. From
this, we can obtain a confidence interval of $v(s,t)$ with following form $(v_{lower}(s,t),v_{upper}(s,t))$
where,
\begin{equation}\label{lower}
  v_{lower}(s,t)=\zeta\left(k+\lambda-\xi\sqrt{2(k+2\lambda)}\right),
\end{equation}
and
\begin{equation}\label{upper}
 v_{upper}(s,t)=\zeta\left(k+\lambda+\xi\sqrt{2(k+2\lambda)}\right),
\end{equation}
with $\xi=F^{-1}_{N(0,1)}\left(1-\frac{\alpha}{2}\right)$ and where $F^{-1}_{N(0,1)}$ is the inverse cumulative normal standard distribution.

\section{\Large Inference on the model }\label{sec3}
In this section, suppose we continuously observe a trajectory of a process in the
interval $[0, T]$, we seek to infer the true value of the parametric vector $\theta$. For this purpose, two methods are presented to estimate CIR parameters: the first estimates the drift parameters $\alpha$ and $\beta$ by the maximum likelihood principle, and the second approximates
the diffusion coefficient $\sigma^2$.

\subsection{\normalsize \textbf{Estimation of drift parameters}}

The SDE (\ref{p3}) can be written in the following vectorial form:
$$dx(t)=A_{t}(x(t))\cdot \theta dt+B_t(x(t))dw(t);\quad 0\leq t\leq T,$$
where $\theta=(\alpha,-\beta)^{\ast},\quad A_{t}(x(t))=(1,x(t))$ and $B_t(x(t))=\sigma \sqrt{x(t)}  $.

The maximum likelihood estimator of the vector $\theta$ is given by (see, for example, \cite{ref7,ref8})
\begin{equation}\label{p6}
\widehat{\theta}_T=S^{-1}_T H_T,
\end{equation}

where $H_T$ is the following $2$-vector:
\begin{equation}\label{p7}
H_T=\int _0^T A_{t}^{\ast}(x(t))\left(B_t(x(t))B_t(x(t))\right)^{-1}dx(t)
\end{equation}

 and $S_T$ is a $2\times 2$-matrix:
\begin{equation}\label{p8}
S_T=\int _0^T A_{t}^{\ast}(x(t))\left(B_t(x(t))B_t(x(t))\right)^{-1}A_{t}(x(t)) dt
\end{equation}
and the asterisk $\ast$ denotes the transpose.

The corresponding vector $H_T$ in Equation(\ref{p7}) in this case leads us to
$$H_T^{\ast}=\frac{1}{\sigma^2}\left(\int _0^T \frac{dx(t)}{x(t)},\int _0^Tdx(t)\right), $$
and the corresponding matrix $S_T$ in Equation(\ref{p8}) in this case leads us to

\begin{equation*}
S_T=\frac{1}{\sigma^2}\left(
\begin{array}{ll}
 T   & \int _0^T x(t)dt\\
\\
\int _0^T x(t)dt  & \int _0^T x(t)^2dt
\end{array}
\right).
\end{equation*}

Using Equation (\ref{p6}) and after some calculation, we obtain the expressions
of the estimators
$$
\begin{array}{ll}
 \widehat{\alpha} &=\frac{\left(\int_0^T x(t)^2dt\right)\left(\int_0^T \frac{dx(t)}{x(t)}\right)-\left(\int_0^T dx(t)\right)\left(\int_0^T x(t)dt\right)}{T\int_0^T x(t)^2dt-\left(\int_0^T x(t)dt\right)^2 }, \\
\\
\widehat{\beta}&=\frac{\left(\int_0^T x(t)dt\right)\left(\int_0^T \frac{dx(t)}{x(t)}\right)-\left(T\int_0^T dx(t)\right)}{T\int_0^T x(t)^2dt-\left(\int_0^T x(t)dt\right)^2 }.
\end{array}
$$
The stochastic integrals in the latter expressions can be transformed into
Riemann–Stieljes integrals by using the Itô formula, hence
$$\int_0^T \frac{dx(t)}{x(t)}=\log\left(\frac{X_T}{x_0}\right)+ \frac{\sigma^2}{2}\int_0^T \frac{dt}{x(t)}.$$
Therefore, the resulting maximum likelihood estimators are
\begin{equation}\label{p9}
\begin{array}{ll}
 \widehat{\alpha} &=\frac{\left(\int_0^T x(t)^2dt\right)\left(\log\left(\frac{x(T)}{x_0}\right)+ \frac{\sigma^2}{2}\int_0^T \frac{dt}{x(t)}\right)-\left(x(T)-x_0\right)\left(\int_0^T x(t)dt\right)}{T\int_0^T x(t)^2dt-\left(\int_0^T x(t)dt\right)^2 }, \\
\\
\widehat{\beta}&=\frac{\left(\int_0^T x(t)dt\right)\left(\log\left(\frac{x(T)}{x_0}\right)+ \frac{\sigma^2}{2}\int_0^T \frac{dt}{x(t)}\right)-T\left(x(T)-x_0\right)}{T\int_0^T x(t)^2dt-\left(\int_0^T x(t)dt\right)^2 }.
\end{array}
\end{equation}

In order to use the above expressions to estimate the parameters, we must have continuous observations. In practice,
continuous sample paths are not usually observed. Rather, the state of the diffusion process is observed at a finite number
of time instances ($0 = t_0 < t_1 < ... < t_n = T )$. In the present case, the likelihood function corresponding to
such data is the product of transition densities has a complicate form and it is very difficult to find
the estimators explicitly. We refer to \cite{ref14, ref15},  an alternative estimation procedure that is frequently utilised for
such data is to use the continuous time maximum likelihood estimators with suitable approximations of the integrals
that appear in the expressions (\ref{p9}); specifically, the Riemann-Stieljes integrals are approximated by means of
the trapezoidal formula.

\subsection{\normalsize \textbf{Approximation of the diffusion coefficient}}

\label{sec032}
In this section, we propose two different ways we used to approximate coefficient $\sigma$, if we assume that it has always
a positive value, are as follows:
\begin{description}
  \item[First method.] The coefficient $\sigma$ can be estimated by using an extension of the procedure proposed by Chesney
and Elliot \cite{ref9}. By applying the Itô formula to the transformation $\sqrt{x(t)}$, we obtain the following equation
  \begin{equation*}
   d\left(\sqrt{x(t)}\right)=\frac{x(t)}{2\sqrt{x(t)}}-\frac{\sigma^2}{8\sqrt{x(t)}}dt.
\end{equation*}
Using the following approximation in the interval $[t-1,t]$:
$d\left(\sqrt{x(t)}\right)=\sqrt{x(t)}-\sqrt{x(t-1)}$ and $dx(t)=x(t)-x(t-1).$ Then,
$$x(t)-x(t-1)-2\sqrt{x(t)}\left(\sqrt{x(t)}-\sqrt{x(t-1)}\right)=\frac{\sigma^2}{4}.$$
The resulting estimator has the following form:
$$\widehat{\sigma}=2\left|\sqrt{x(t)}-\sqrt{x(t-1)}\right|.$$

For $n $ observations of one trajectory of the process, the resulting estimator has the following expression:
\begin{equation}\label{ref10}
\widehat{\sigma}=\frac{2}{n-1}\sum^n_{t=2} \left|\sqrt{x(t)}-\sqrt{x(t-1)}\right|.
\end{equation}
  \item[Second method.] The coefficient $\sigma$ can be estimated by using an extension of the procedure proposed by A. Katsamaki and C. H. Skiadas \cite{ref18}. From the stochastic differential equation \ref{p3}, we get
      \begin{align*}
        dx(t)&=\left(\alpha-\beta x(t)\right)dt+\sigma\sqrt{x(t)}dw(t) \\
        & \Longleftrightarrow \left(dx(t)\right)^2=\left(\sigma\sqrt{x(t)}\right)^2dt.
      \end{align*}

Then, $\left(\frac{dx(t)}{\sqrt{x(t)}}\right)^2=\sigma^2dt.$
 Considering that $dx(t)=x(t)-x(t-1)$ a second approximation for $\sigma$, is:
$$\widehat{\sigma}=\left|\frac{x(t)-x(t-1)}{\sqrt{x(t)}}\right|.$$
For $n $ observations of one trajectory of the process, the resulting estimator has the following expression:
\begin{equation}\label{ref11}
\widehat{\sigma}=\frac{1}{n-1}\sum^n_{t=2} \left|\frac{x(t)-x(t-1)}{\sqrt{x(t)}}\right|.
\end{equation}
\end{description}

\begin{remark}
{\normalfont By using Zehna's theorem, the estimated conditional trend function
(ECTF) of the CIR is obtained by replacing the parameters in expression by Eqs. \ref{ref10} and \ref{p9} or \ref{ref11} and \ref{p9}, and thus the ECTF is given by
\begin{equation}\label{ECTF}
 \widehat{E}(x(t)|x(s)=x_{s})= x_{s} e^{\widehat{\beta} (s-t)}+\frac{\widehat{\alpha}}{\widehat{\beta}}\left(1-e^{\widehat{\beta} (s-t)}\right).
\end{equation}}
\end{remark}

\section{\Large Simulated sample paths of the process}\label{sec5}
\subsection{\normalsize \textbf{Simulated sample paths of the process}}\label{sec024}
\begin{figure}[thbp]
\centering
  \includegraphics[width=7cm]{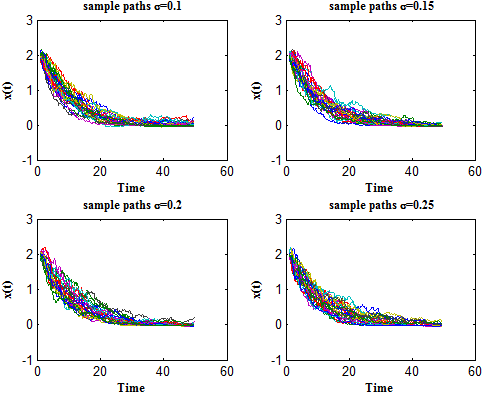}
  \caption{\textsf{Simulated sample paths for the SCDP for several values of $\sigma$($x_0=2,\beta=0.1, \alpha=0.2$).}}\label{fig1}
\end{figure}

\begin{figure}[thbp]
 \centering
  \includegraphics[width=7cm]{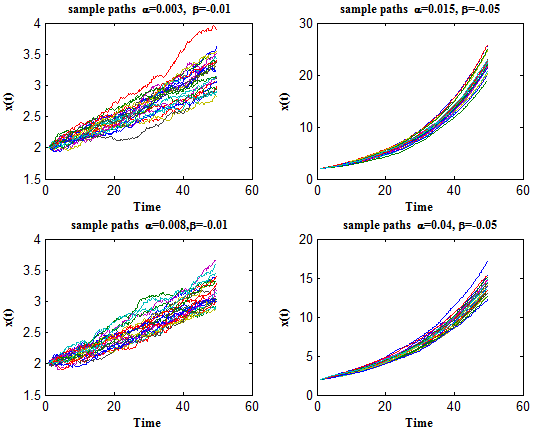}
  \caption{\textsf{Simulated sample paths for the SCDP for several values of $\alpha$ and $\beta$ ($x_0=2,\sigma=0.02$).}}\label{fig2}
\end{figure}

In this section, we present some simulated sample paths for the CIR. By using procedure proposed by P. Kloeden, E. Platen \cite{ref3}, such as Taylor’s algorithm to the order of 1.5 in time intervals of length h, from which in the case of the CIR diffusion process, we have
\begin{equation}\label{algorithm}
 \begin{split}
    x_{n+1}&=\frac{\sigma}{2\sqrt{x_n}}\left\{\left(2x_n+ \alpha h-\beta h x_n-\frac{\sigma^2 h}{4}\right)\Delta W\right.\\
    &+\left.\left(\frac{\sigma^2 }{4}-\beta x_n-\alpha\right)\Delta Z\right\}
    +\left(1-\beta h+\frac{ \beta^2 h^2}{2}\right)x_n\\
    &+\frac{\sigma^2}{4}(\Delta W)^2
    +\alpha h -\frac{\sigma^2 h}{4}-\frac{\alpha \beta h^2}{2};  x(t_0)=x_0.
  \end{split}
\end{equation}
where $\Delta W=\sqrt{h}U_1$ and $\Delta Z=\frac{h^{\frac{3}{2}}}{2}\left(U_1+\frac{U_2}{\sqrt{3}}\right)$ , with $U_1$ and $U_2$ being two standard normal distribution independent random variables, and where $h$ is the discretization step.
Figure \ref{fig1} and Figure \ref{fig2} shows the some simulated sample paths for the CIR for several values of $\alpha$, $\beta$ and $\sigma$.

\subsection{\normalsize \textbf{Simulation Examples}}

In this section we present several examples in order to validate the estimation procedure previously developed in section (\ref{sec3}). To this end, we have considered an example in which $N=25$ sample paths
have been simulated. Each trajectory has been simulated with $t_i=t_{i-1}+(i-1)h$; for $i=2,...,N$ starting at $t_1=0$, taking the step size $h=1$ and $x_0=2$. The statistical  methodology can be applied in the following phases: First, use the all data to estimate the parameters $\beta$ and $\alpha$ of the process, using the expression  \ref{p9} and to approximate $\sigma^2$ by the approximation \ref{ref10} and \ref{ref11}. Moreover, obtain the corresponding ETF and ECTF values given by the expression \ref{TF} and \ref{CTF}.
 To illustrate the performance of procedure, the results according to the one-step-ahead mean absolute error (MAE), the root mean square error (RMSE) and the mean absolute percentage error (MAPE), given by Table \ref{tab0}. According to Lewis \cite{ref16}, we deduce the accuracy of the forecast can be judged from the MAPE result Table \ref{tab10}.

 \begin{table}[H]
\centering
\footnotesize
\caption{The one-step-ahead mean absolute error, the root mean square error, and mean absolute percentage error.}
\hfill{}
\begin{tabular}{l}
  \hline
  MAE= $\frac{1}{N}\sum\limits_{i=1}^N |x(t_i)-\hat{x}(t_i)|$,\\
 RMSE= $\sqrt{\frac{1}{N}\sum\limits_{i=1}^N(x(t_i)-\hat{x}(t_i))^2}$,\\
  MAPE=$\frac{1}{N}\sum\limits_{i=1}^N\frac{|x(t_i)-\hat{x}(t_i)|}{x(t_i)}\times 100$.\\
\hline
\end{tabular}
\hfill{}
\label{tab0}
\end{table}

\begin{table}[H]
\centering
\footnotesize
\caption{Interpretation of typical MAPE values.}
\hfill{}
 \begin{tabular}{ll}
  MAPE &  Interpretation \\
  \hline
 $<$10 & Highly accurate forecasting \\
  10 \:30& Good forecasting \\
 30 \:50 & Reasonable forecasting \\
 $>$ 50& Inaccurate forecasting \\
\hline
\end{tabular}
\hfill{}
\label{tab10}
\end{table}

\begin{figure}[H]
\centering
  \includegraphics[width=7cm]{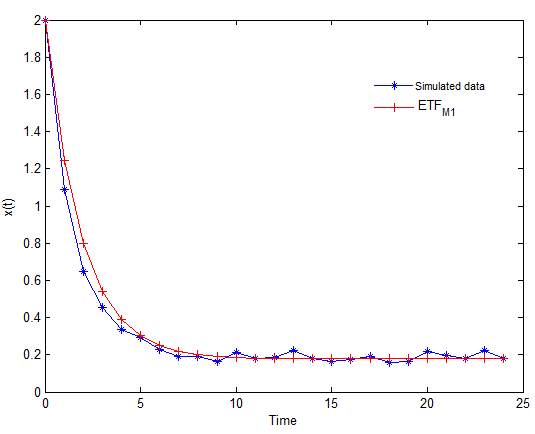}
  \caption{Simulated data versus estimated trend function (ETF$_{M1}$).}\label{fig10}
\end{figure}%

\begin{figure}[H]
\centering
  \includegraphics[width=7cm]{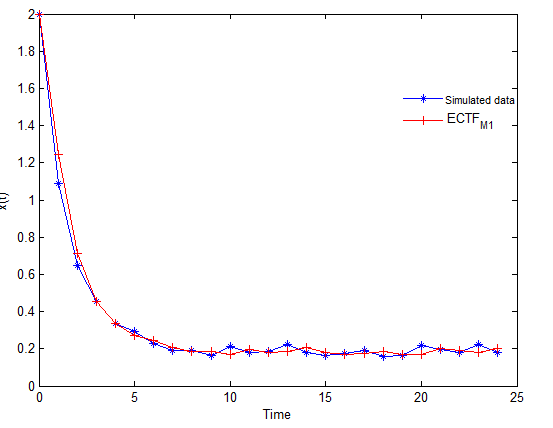}
  \caption{Simulated data versus estimated conditional trend function (ECTF$_{M1}$).}\label{fig11}
\end{figure}

\begin{figure}[H]
\centering
  \includegraphics[width=7cm]{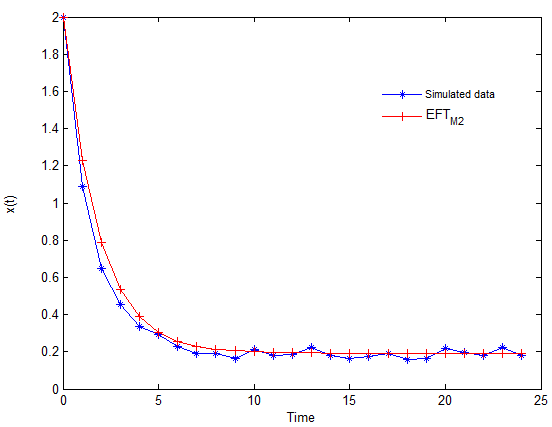}
  \caption{Simulated data versus estimated trend function (ETF$_{M2}$).}\label{fig12}
\end{figure}

\begin{figure}[H]
\centering
  \includegraphics[width=7cm]{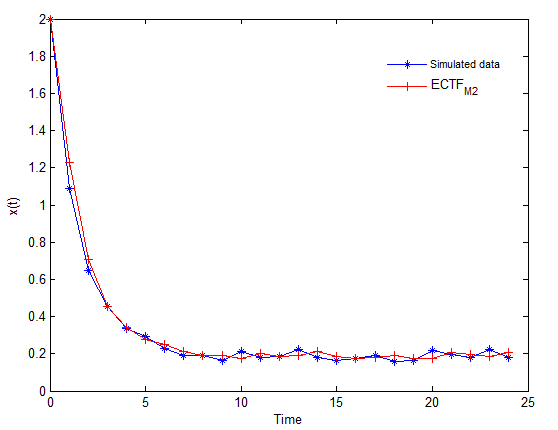}
  \caption{Simulated data versus estimated conditional trend function (ECTF$_{M2}$).}\label{fig13}
\end{figure}

\begin{table*}[ht]
\centering
\footnotesize
\caption{Estimation of the parameters of the process using all data.}
\hfill{}
 \begin{tabular}{ccc}
\hline
  Parameters & $\widehat{\sigma}=\frac{2}{n-1}\sum^n\limits_{t=2} \left|\sqrt{x(t)}-\sqrt{x(t-1)}\right| $ & $\widehat{\sigma}=\frac{1}{n-1}\sum^n\limits_{t=2} \left|\frac{x(t)-x(t-1)}{\sqrt{x(t)}}\right|$  \\
  \hline
    $\alpha$ &   0.095418335305905&  0.106759504575717\\
   $\beta$ &     0.537546048354900 & 0.555288145235607 \\
  $\sigma$ &     0.125145131849032 &  0.135845262598210\\
\hline
\end{tabular}
\hfill{}
\label{tab2}
\end{table*}
Table \ref{tab2} shows the estimation of the parameters of the process using the expressions \ref{p9}, \ref{ref10} and \ref{ref11}.
Table \ref{tab1} shows the results for the ETF$_{M1}$, ECTF$_{M1}$, ETF$_{M2}$ and ECTF$_{M2}$ of the process. Table \ref{tab3} shows the goodness of fit of the process. The accuracy of the forecast can be judged from the MAPE result is less than 10\%, showing the forecast to be highly accurate for the first approximation of $\sigma$ and the MAPE result is between 10\% and 30\%, showing the forecast to be good forecasting for the second approximation of $\sigma$. The performance of the CIR for the forecasting using the trend function and the conditional trend function for the data is illustrated in Figure \ref{fig10}, Figure \ref{fig11}, Figure \ref{fig12} and Figure \ref{fig13}.
 \begin{table*}[ht]
\centering
\footnotesize
\caption{Goodness of fit of the process.}
\hfill{}
\begin{tabular}{ccc}
  \hline
 & $\widehat{\sigma}=\frac{2}{n-1}\sum^n\limits_{t=2} \left|\sqrt{x(t)}-\sqrt{x(t-1)}\right| $ & $\widehat{\sigma}=\frac{1}{n-1}\sum^n\limits_{t=2} \left|\frac{x(t)-x(t-1)}{\sqrt{x(t)}}\right|$  \\
  \hline
MAE & 0.0314080&   0.0331183\\
  RMSE & 0.0514486& 0.0485530 \\
MAPE &   9.74\%& 11.55\% \\
\hline
\end{tabular}
\hfill{}
\label{tab3}
\end{table*}

\begin{table}[H]
\centering
\footnotesize
\caption{Simulated data, showing ETF$_{M1}$, ECTF$_{M1}$,  ETF$_{M2}$ and  ECTF$_{M2}$ using the model.}
\hfill{}
\begin{tabular}{cccccc}
   \hline
 i  &  $x(t_i)$  & ETF$_{M1}$ & ECTF$_{M1}$ &ETF$_{M2}$ &ECTF$_{M2}$ \\
  \hline
    1   & 2.00000 & 2.00000 &2.00000  &   2.00000 &2.00000  \\
   2   &   1.09087&  1.24217 & 1.24217 & 1.22973&1.22973 \\
   3   &   0.64860& 0.79946 & 0.71107 &0.78767 & 0.70797\\
   4   &   0.45218 &  0.54084 & 0.45271 & 0.53397 & 0.45415 \\
   5   &   0.33668&  0.38976& 0.33796  & 0.38837& 0.34142\\
   6   & 0.29343&0.30150& 0.27049 &  0.30481& 0.27514\\
   7   & 0.23103& 0.24994 &0.24522 &  0.25685& 0.25032\\
   8   &   0.19187 &0.21982&0.20877 & 0.22933& 0.21451\\
   9   &   0.19174&0.20222& 0.18589 & 0.21353& 0.19203\\
   10  & 0.16141 & 0.19195& 0.18582&  0.20447&  0.19196\\
   11  & 0.21271 & 0.18594&0.16811&   0.19926& 0.17455 \\
   12  &0.17947& 0.18243& 0.19807& 0.19628 &0.20399 \\
   13  &0.18705&  0.18038&  0.17865&  0.19456&  0.18492 \\
   14  &0.22376&  0.17919&0.18308&0.19358&  0.18926\\
   15  & 0.17811 & 0.17849& 0.20452 &  0.19302 & 0.21033\\
   16  & 0.16636& 0.17808 &0.17785& 0.19269&0.18414\\
   17  &0.17499&0.17784 & 0.17099&   0.19251 &0.17739\\
   18  &0.19105& 0.17770& 0.17603&  0.19240 &0.18234\\
   19  &0.16005 &0.17762 &0.18541& 0.19234 & 0.19156\\
   20  &0.16561& 0.17757&0.167309&  0.19230 & 0.17377\\
   21  & 0.21905 & 0.17754&0.17055& 0.19229   &0.17696\\
  22  &0.19761& 0.17753& 0.20177&   0.19228& 0.20763\\
  23  & 0.17825 &  0.17752& 0.18925 &  0.19227&0.19533\\
  24  &  0.22156 & 0.17751& 0.17794&   0.19226 &0.18421\\
  25  & 0.17769 & 0.17751& 0.20324&   0.19226&0.20907\\
\hline
\end{tabular}
\hfill{}
\label{tab1}
\end{table}

\section{\Large Conclusions} \label{sec5}
In this study, we introduced the CIR diffusion process. Its distribution and main characteristics were analyzed, and its trend function as well as its conditional trend function was found by a parametrization of the CIR process.

The inferential study is carried on the basis of continuous sampling via the maximum likelihood method. Since a maximum
likelihood estimators with suitable approximations of the integrals that appear in the expressions; specifically, the Riemann-Stieljes integrals are approximated by means of the trapezoidal formula. The diffusion coefficient is approximate by two different methods.

Finally, the variable under study, could be generalized in the future studies.
\section*{Acknowledgements}
The authors are very grateful to Editor and referees for constructive comments and suggestions. This research has been funded by LAMSAD from $''$Fonds propres de l'Universit$\acute{\mbox{e}}$ Hassan First of Settat, (Morocco)$''$.

\noindent\hrulefill
\renewcommand\refname{REFERENCES}

\end{document}